\documentclass[aps,twocolumn,prl,superscriptaddress,longbibliography,reprint]{revtex4-2}
\usepackage{amsfonts}
\usepackage{mathrsfs}
\usepackage{amsmath}
\usepackage{color}
\usepackage{graphicx}
\usepackage{bm}
\usepackage{amssymb}
\usepackage{xspace}
\usepackage{epstopdf}
\usepackage{longtable}
\usepackage{multirow}
\usepackage{booktabs}
\usepackage{textcomp}
\usepackage[title]{appendix}
\usepackage{float}
\usepackage{dsfont}
\usepackage[colorlinks=true, letterpaper=true, pdfstartview=FitV, linkcolor=blue, citecolor=blue, urlcolor=blue]{hyperref}
\usepackage[]{changes}

\providecommand{\U}[1]{\protect\rule{.1in}{.1in}}

\begin{document}

\title{Nonperturbative Magnetic Orbital Hall Effect in Altermagnets}

\begin{abstract}

Recent studies on altermagnets have focused considerable attention on nonrelativistic effects that persist in the absence of spin-orbit coupling (SOC). As a result, the relative importance of various phenomena in altermagnets has commonly been judged by their dependence on SOC. Here, we challenge this
common wisdom by uncovering the magnetic orbital Hall effect, which is nonperturbative in SOC strength.
We establish the symmetry properties of this effect, demonstrating that it is strictly forbidden in conventional collinear antiferromagnets yet universally allowed in all ten spin-Laue classes of collinear altermagnets. Counterintuitively, although SOC-induced, it reaches giant magnitudes in altermagnets --- comparable to or even exceeding the nonrelativistic spin Hall effect.
Moreover, altermagnetic symmetry enables unconventional collinear-polarized orbital currents, allowing field-free manipulation of perpendicular magnetization.
Our first-principles calculations predict strong room-temperature responses in the experimentally established altermagnets CrSb and FeSb$_2$. These findings reveal the previously overlooked potential of altermagnetic orbitronics and broaden the horizons for altermagnets in high-performance magnetic memory applications.
\end{abstract}

\author{Xukun Feng}
\affiliation{Interdisciplinary Center for Theoretical Physics and Information Sciences (ICTPIS), Fudan University, Shanghai 200433, China}

\author{Jin Cao}
\affiliation{Research Laboratory for Quantum Materials, Department of Applied Physics, The Hong Kong Polytechnic University, Kowloon, Hong Kong, China}

\author{Lay Kee Ang}
\affiliation{Science, Mathematics and Technology, Singapore University of Technology and Design, Singapore 487372, Singapore}

\author{Shengyuan A. Yang}
\email{shengyuan.yang@polyu.edu.hk}
\affiliation{Research Laboratory for Quantum Materials, Department of Applied Physics, The Hong Kong Polytechnic University, Kowloon, Hong Kong, China}

\author{Cong Xiao}
\email{congxiao@fudan.edu.cn}
\affiliation{Interdisciplinary Center for Theoretical Physics and Information Sciences (ICTPIS), Fudan University, Shanghai 200433, China}

\author{X. C. Xie}
\affiliation{Interdisciplinary Center for Theoretical Physics and Information Sciences (ICTPIS), Fudan University, Shanghai 200433, China}
\affiliation{International Center for Quantum Materials, School of Physics, Peking University, Beijing 100871, China}

\maketitle

Orbital Hall effect --- the generation of a transverse orbital angular momentum current in response to an applied electric field --- has emerged as a vibrant area of research \cite{bernevig2005orbitronics,Guo2005,kontani2009giant,go2018intrinsic,Bhowal2020,sala2022giant,Salemi2022PRM,Manchon2022OHE,Rappoport2023orbital,Mertig2024OHE}. It has been intensively studied in nonmagnetic systems with weak spin-orbit coupling (SOC), such as Ti and Cr \cite{choi2023observation,Kawakami2023OHE-Cr}. In these systems, unlike its spin counterpart (i.e., the spin Hall effect), the orbital response does not require SOC, enabling orbital currents that can be orders of magnitude stronger than spin Hall effect \cite{jo2018gigantic,choi2023observation,Kawakami2023OHE-Cr}. When injected into an
adjacent magnetic layer, such currents can efficiently drive magnetic switching, offering significant potential for magnetic memory applications  \cite{Zheng2020SOT-Zr,go2020orbital,Ding2020OT-CuO,Otani2021OT,lee2021efficient,lee2021orbital,choi2023observation,zheng2024effective,ding2024orbital,Jiang2024OT-PMA-Zr,Klaui2025OT-PMA-Ru,zhang2025orbital,Wang2025Orbital-SRO}.

In \emph{magnetic} systems, a distinct variant known as the magnetic orbital Hall effect (MOHE) arises~\cite{salemi2022theory}. While the conventional orbital Hall effect in nonmagnetic materials is even under time-reversal ($\mathcal T$) operation, MOHE is odd under $\mathcal T$, meaning that its sign reverses upon flipping the magnetic order. This property provides an additional handle for controlling orbital currents, making MOHE particularly attractive for spintronic and orbitronic devices.

To date, the study on MOHE is rare~\cite{salemi2022theory}, and has focused on ferromagnetic materials. For practical applications, however, antiferromagnets are preferable as orbital current sources, because they produce minimal stray fields and reduced interference with neighboring magnetic layers. Nevertheless, MOHE in antiferromagnets remains unexplored. This absence stems from two main challenges. (i) One can show that MOHE is forbidden
in \emph{conventional} antiferromagnets with $\mathcal{PT}$ or $t_{1/2}\mathcal T$ symmetry ($\mathcal{P}$ denotes inversion and $t_{1/2}$ a fractional translation).
It is not yet clear what kind of antiferromagnets can support MOHE.
(ii) In collinear antiferromagnets, MOHE must require SOC, leading to the widespread expectation of only weak responses.
However, is this expectation really the case? Is it possible to realize strong MOHE in antiferromagnets? And furthermore, what is the guideline for finding such materials? These key questions remain to be answered.

In this work, we address these outstanding questions. We show that although MOHE is forbidden in conventional antiferromagnets, it is universally allowed in altermagnets --- a newly recognized class of collinear antiferromagnets that has become a focus of current research~\cite{Libor2022,smejkal2022prx,wu2007Fermi,Naka2019spincurrent,hayami2019momentum,smejkal2020crystal,Yuan2020Giant,shao2021,ma2021multifunctional,SST2021,Libor2022TMR,Shao2023Neel,Bai2024altermagnetism,Yu2024SLC,song2025altermagnets}. From symmetry analysis, we demonstrate that MOHE exists in all ten spin-group classes that encompass known altermagnets, establishing it as a transport signature that distinguishes altermagnets from conventional antiferromagnets. Our analysis also gives explicit results of how MOHE conductivity scales with spin-conserve SOC and spin-flip SOC~\cite{zhang2011PRL,Guo2014,Liu2025PRX,2025Agterberg-quasisymmetry} in the perturbative regime. Importantly, we show that although MOHE here must require SOC, its dependence on SOC can be \emph{nonperturbative}, meaning that the MOHE magnitude is not necessarily small; instead, can be remarkably large even under weak SOC. This happens when the Fermi level lies near a small SOC-induced gap. These predictions are confirmed through an effective lattice model and first-principles calculations on realistic altermagnetic materials CrSb and FeSb$_2$.
Particularly, in these materials, MOHE can dominate over the spin Hall effect (even for spin Hall response with nonrelativistic origin), reaching significant values comparable to or exceeding the previously reported orbital Hall responses. Moreover, MOHE in altermagnets enables the generation of collinearly-polarized orbital current (CPOC), a highly desirable feature for information devices with improved storage density and endurance \cite{kurebayashi2017view,Ryu2020SOT,Yang2023briefing}. Our findings position altermagnets as promising orbital current sources and establish MOHE as a new hallmark of this emerging class of materials, opening a promising route toward next-generation information technologies.

\textcolor{blue}{\textit{Symmetry analysis.}}
The orbital current response is characterized by a rank-3 pseudotensor $\sigma_{bc}^{L_a}$ defined by
\begin{eqnarray}
  j_{b}^{L_a} = \sigma_{bc}^{L_a} E_c,
\end{eqnarray}
where $j_{b}^{L_a}$ is the orbital angular momentum current density flowing in direction $b$ and polarized in direction $a$, $E$ is the applied electric field, roman indices denote Cartesian components in real space, roman indices $(=x,y,z)$ denote the Cartesian components, and repeated indices are summed over.

\begin{figure}
	\includegraphics[width=8.5cm]{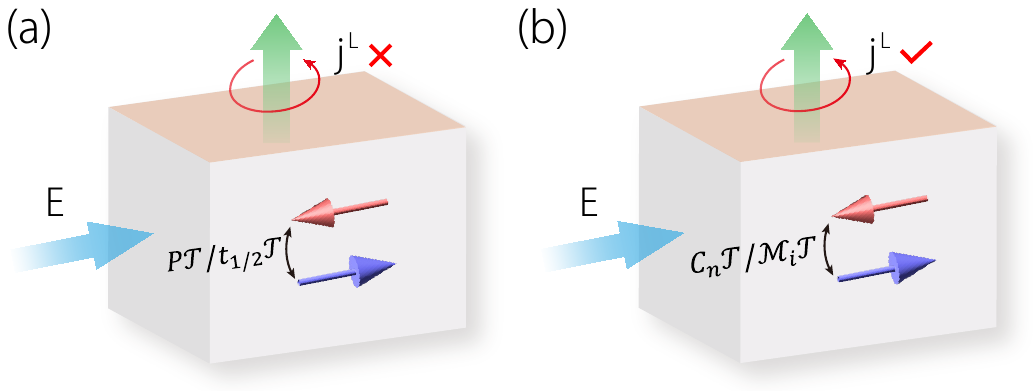}
	\caption{Illustration of MOHE in antiferromagnets. (a) The effect is forbidden in conventional antiferromagnets, where the magnetic sublattices (denoted by red and blue arrows) are connected by $\mathcal{PT}$ or $t_{1/2}\mathcal T$ symmetry.  (b) It is generally permitted in altermagnets with the magnetic sublattices connected by $C_n\mathcal T$ or $M_i\mathcal T$ symmetries ($C$ and $M$ are rotation and mirror operations, respectively).
		\label{figconcept}}
\end{figure}

First of all, let us see why the MOHE is forbidden in conventional antiferromagnets. By definition, MOHE is odd under $\mathcal T$. For antiferromagnets with $t_{1/2}\mathcal T$ symmetry, its magnetic point group contains $\mathcal T$, allowing only
$\mathcal T$-even responses, so MOHE is forbidden. As for systems with $P\mathcal T$ symmetry, one notes that both $j^{L}$ and $E$ are odd under $P\mathcal T$, which also forbids a $\mathcal T$-odd response [see Fig.~\ref{figconcept}(a)].
In comparison, $t_{1/2}\mathcal T$ and $P\mathcal T$ are broken in altermagnets, where the two magnetic sublattices are connected by certain mirror or rotational symmetry. We shall see altermagnets generally permit a nonzero MOHE [see Fig.~\ref{figconcept}(b)].

It was found that altermagnets can host spin Hall effect in the absence of SOC \cite{SST2021}. However, this is not the case for MOHE. To see this, we note that without SOC, an altermagnet is described by spin groups~\cite{Libor2022,smejkal2022prx,Liu2022spingroup}. The spin-only group always contains a $C_2'=C_2 \mathcal T$ symmetry, where $C_2$ is a reversal of spin, having no action on orbitals. It follows that $C_2'$ just acts like an effective time-reversal symmetry for orbital transport, thus forbidding MOHE. This demonstrates that nonzero MOHE in altermagnets must require SOC.

\begin{table*}
	\caption{\label{tableSOC}Symmetry-allowed \emph{collinearly-polarized} components of MOHE and spin Hall conductivities for the 10 spin Laue groups relevant to collinear altermagnets. `$\times$' represents the component is forbidden. Here, we take $\bm N\|\hat z$. The symbols with blue (red) color correspond to the responses with polarization parallel (perpendicular) to the N\'eel vector.}
    \renewcommand{\arraystretch}{1.3}
	\begin{ruledtabular}
		\begin{tabular}{ccccc}		
			Spin Laue Group & Anisotropy & Nonrelativistic limit& Spin-conserve SOC & Spin-flip SOC  \\
			\hline
            \noalign{\vskip 1pt}
            ${}^{2}m^{2}m^{1}m$ & $d$-wave & $\times$ & $\times$ & $\times$ \\
            ${}^{2}4/^{1}m$ & $d$-wave & $\times$& $\color{red}\ell_{xz,z}^{L_x,3}$, $\color{red}\ell_{yz,z}^{L_y,3}$ & $\color{red}q_{xz,xy}^{L_x,12}$, $\color{red}q_{yz,xy}^{L_y,12}$,
            $\color{red}\ell_{xz,y}^{S_x,2}$, $\color{red}\ell_{yz,x}^{S_y,1}$ \\
            ${}^{2}4/^{1}m^{2}m^{1}m$ & $d$-wave & $\times$ & $\color{red}\ell_{xz,z}^{L_x,3}$,
            $\color{red}\ell_{yz,z}^{L_y,3}$ & $\color{red}\ell_{xz,y}^{S_x,2}$, $\color{red}\ell_{yz,x}^{S_y,1}$, $\color{red}q_{xz,xy}^{L_x,12}$, $\color{red}q_{yz,xy}^{L_y,12}$ \\
            ${}^{1}4/^{1}m^{2}m^{2}m$ & $g$-wave & $\times$ & $\times$ & $\times$ \\
            ${}^{1}6/^{1}m^{2}m^{2}m$ & $i$-wave & $\times$ & $\times$ & $\times$ \\
            ${}^{2}2/^{2}m$ & $d$-wave & $\color{blue}\sigma_{zy}^{S_z,(0)}$&
            $\color{blue}\ell_{zy,z}^{L_z,3}$,
            $\color{blue}q_{zy,zz}^{S_z,33}$;
            $\color{red}\ell_{xy,z}^{L_x,3}$ &
            $\color{blue}q_{zy,xy}^{L_z,12}$, $\color{blue}q_{zy,xx}^{S_z,11}$, $\color{blue}q_{zy,yy}^{S_z,22}$; $\color{red}q_{xy,xy}^{L_x,12}$,
            $\color{red}\ell_{xy,y}^{S_x,2}$ \\
            ${}^{1}\overline{3}{}^{2}m$ & $g$-wave & $\times$& $\color{red}\ell_{yx,z}^{L_y,3}$ & $\color{red}q_{yx,xy}^{L_y,12}$,
            $\color{red}\ell_{yx,x}^{S_y,1}$ \\
            ${}^{2}6/^{2}m$ & $g$-wave & $\times$& $\color{red}\ell_{xy,z}^{L_x,3}$, $\color{red}\ell_{yx,z}^{L_y,3}$ & $\color{red}q_{xy,xy}^{L_x,12}$, $\color{red}q_{yx,xy}^{L_y,12}$,
            $\color{red}\ell_{xy,y}^{S_x,2}$, $\color{red}\ell_{yx,x}^{S_y,1}$ \\
            ${}^{2}6/^{2}m^{2}m^{1}m$ & $g$-wave & $\times$& $\color{red}\ell_{xy,z}^{L_x,3}$ & $\color{red}q_{xy,xy}^{L_x,12}$,
            $\color{red}\ell_{xy,y}^{S_x,2}$ \\
            ${}^{1}m^{1}\overline{3}{}^{2}m$ & $i$-wave & $\times$ & $\color{red}\ell_{xz,z}^{L_x,3}$,
            $\color{red}\ell_{yz,z}^{L_y,3}$ &  $\color{red}q_{xz,xy}^{L_x,12}$, $\color{red}q_{yz,xy}^{L_y,12}$,
            $\color{red}\ell_{xz,y}^{S_x,2}$, $\color{red}\ell_{yz,x}^{S_y,1}$\\
		\end{tabular}
	\end{ruledtabular}
\end{table*}


Next, we analyze the symmetry-allowed structure of the MOHE conductivity tensor by expanding  $\sigma_{bc}^{L_a}$ in powers of SOC and constraining the expansion with spin point group. This approach allows us to find how MOHE scales with different kinds of SOCs in the perturbative regime, i.e., when SOC is weak compared to other energy scales in the system. Following Ref.~\cite{Liu2025PRX}, we introduce spin-orbit vectors $\bm \zeta^\alpha$, which characterizes the form of SOC via
\begin{eqnarray}
  H_\text{SOC}=\lambda  \hat s_\alpha(\bm \zeta^\alpha\cdot\hat{\bm O}),
\end{eqnarray}
where $\hat s_\alpha$ ($\alpha=1,2,3$ labels spin space directions) are the three spin operators, $\hat{\bm O}$ is the orbital operator acting on real space, and  $\lambda$ is the SOC strength. The $\bm \zeta^\alpha$ vectors help to distinguish spin-conserve and spin-flip SOC terms (SOC terms with $\bm \zeta^3$ are spin-conserve, whereas those with $\bm \zeta^1$ and $\bm \zeta^2$ are spin-flip, as detailed in Supplemental Material \cite{supp}).

The perturbative expansion of $\sigma_{bc}^{L_a}$ takes the form of
\begin{align}
\sigma_{bc}^{L_a}
&= \ell_{bc,d}^{L_a,\alpha} \zeta_d^{\alpha}
   + q_{bc,de}^{L_a,\alpha\beta} \zeta_d^{\alpha} \zeta_e^{\beta} + \cdots.\label{expansion}
\end{align}
Here, $\ell$'s and $q$'s are expansion coefficients for linear and quadratic terms, respectively. In collinear altermagnets, as discussed, there is no zeroth order term for MOHE. And another important common feature imposed by the spin-only group of collinear magnets is: the linear terms contain only spin-conserve SOC, while quadratic terms contain only spin-flip SOC~\cite{supp}. Therefore, up to second order of SOC, the spin-only group enforces
\begin{eqnarray}\label{44}
  \sigma_{bc}^{L_a}
= \ell_{bc,d}^{L_a,3} \zeta_d^{3}
   + q_{bc,de}^{L_a,12} (\zeta_d^{1} \zeta_e^{2}- \zeta_d^{2} \zeta_e^{1}).
\end{eqnarray}

Consider the ten spin Laue groups that cover the existing altermagnetic materials. Their allowed coefficients in (\ref{44}) can be screened out. Here, we focus on the components that produce CPOC, i.e., with $a=b$.
Table~\ref{tableSOC} shows the results for the case $\bm N\|\hat z$ (results for $\bm N$ along $x$ and $y$ directions are presented in Supplemental Material \cite{supp}). For comparison, we also list the allowed coefficients for magnetic spin Hall effect in the table. One has the following observations. First, one sees that CPOC from MOHE can be generated in most (7 out of 10) classes. For the remaining three classes, although they do not support CPOC (for  $\bm N\|\hat z$), other non-CPOC components of $\sigma_{bc}^{L_a}$ are still allowed, demonstrating that MOHE is universally present in altermagnets. Second,
regarding the previously studied spin Hall effect without SOC, its presence is actually quite restricted, only in the ${}^{2}2/^{2}m$ class here. This indicates that in most altermagnets, actually both magnetic spin Hall effect and MOHE are  induced by SOC, suggesting that MOHE is not necessarily weaker than spin response even in the perturbative regime.
Third, for the three classes ${}^{2}m^{2}m^{1}m$, ${}^{1}4/^{1}m^{2}m^{2}m$ and ${}^{1}6/^{1}m^{2}m^{2}m$, as shown in Supplemental Material \cite{supp}, CPOC can still be generated when the N\'{e}el vector is along the $x$ or $y$ direction. Finally, we mention that since MOHE requires SOC, its symmetry-allowed components may also be analyzed by using magnetic groups. This has also been done \cite{supp}, and the results are consistent with those in Table~\ref{tableSOC}.


\begin{figure}
	\includegraphics[width=8.5cm]{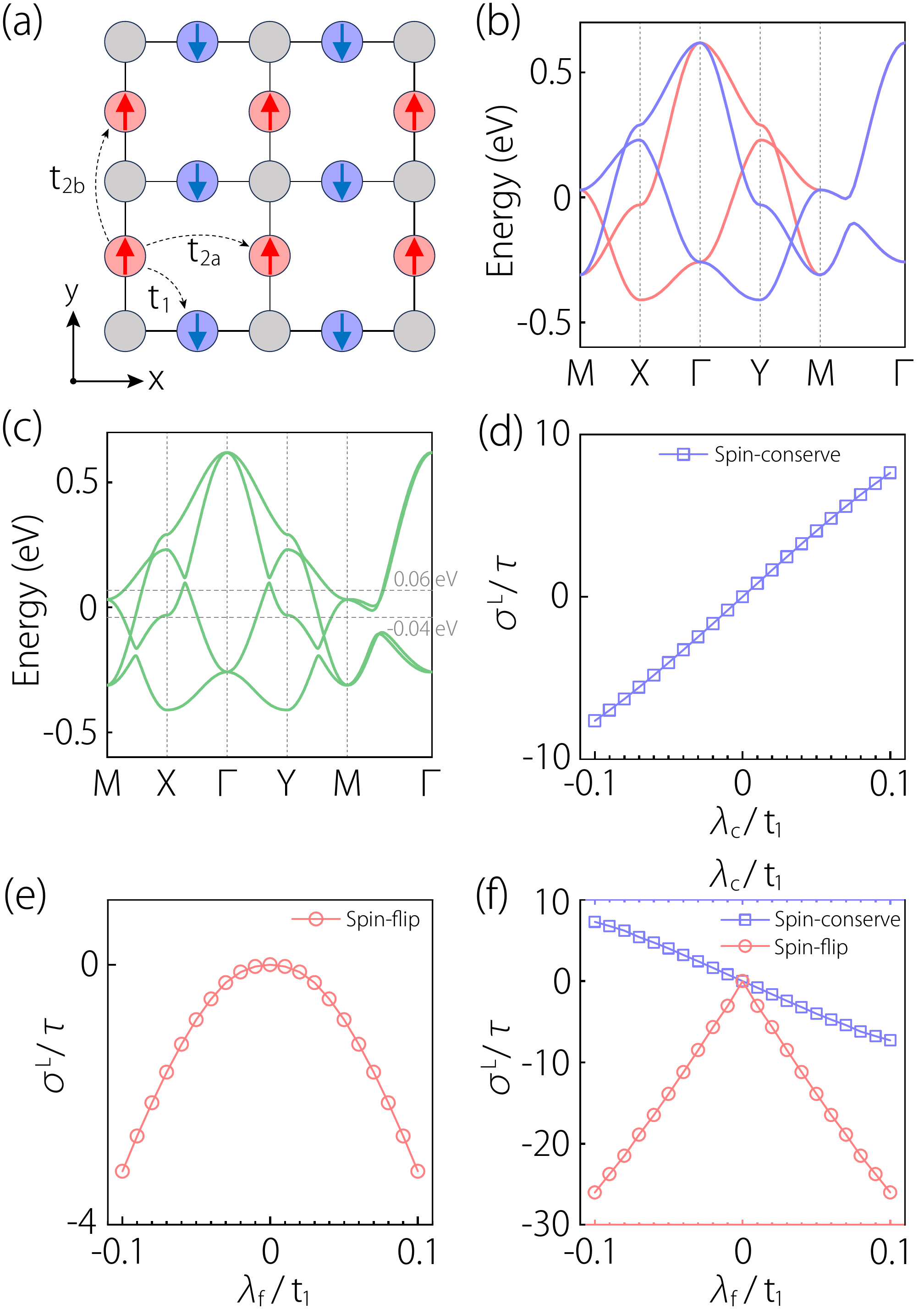}
	\caption{(a) Illustration of the 2D $d$-wave altermagnetic model in (\ref{model}). (b) Calculated non-relativistic band structure. Spin-up and spin-down bands are denoted by red and blue colors. (c) Band structure with SOC strength $\lambda_{c}=\lambda_f=0.1t_1$. (d) $\sigma^L/\tau$ versus spin-conserve SOC strength $\lambda_c$. The spin-flip SOC is turned off. (e) Same as (d), but with the roles of spin-flip and spin-conserve SOC interchanged. In (d) and (e), the chemical potential is set to $\mu=-0.04$ eV. (f) The result at $\mu=0.06$ eV near an SOC-gap. In the calculation, we take $4t_1=0.2$, $2t_2=-0.08$, $2t_d=-0.04$, and $J=0.11$. The unit of $\sigma^L /\tau$ is $(\hbar/e)\,\Omega^{-1}\mathrm{cm}^{-1}\text{fs}^{-1}$.
		\label{figmodel}}
\end{figure}

\textcolor{blue}{\textit{Nonperturbative enhancement of MOHE.}}
To illustrate the features discussed above, we study a minimal model for a two-dimensional (2D) $d$-wave altermagnet on a square lattice [see Fig.~\ref{figmodel}(a)], which belongs to the spin Laue group ${}^{2}4/^{1}m^{2}m^{1}m$~\cite{model2024PRB,antonenko2025mirror}:
\begin{equation}
    \mathcal{H} = \mathcal{H}_0 + \mathcal{H}_{\text{SOC}}.\label{model}
\end{equation}
Here,
\begin{eqnarray}
  \mathcal{H}_0&=&4t_1\cos(k_x/2)\cos(k_y/2)\tau_x+2t_2\left[\cos(k_x)+\cos(k_y)\right]\tau_0 \nonumber \\& &\qquad+2t_s\left[\cos(k_x)-\cos(k_y)\right]\tau_z+J\tau_z\bm{N}\cdot\bm{\sigma},
\end{eqnarray}
where $\tau$'s and $\sigma$'s are Pauli matrices denoting respectively the sublattice and spin degrees of freedom, $t$'s are hopping parameters with $t_2=\frac{1}{2}(t_{2a}+t_{2b})$ and $t_s=\frac{1}{2}(t_{2a}-t_{2b})$ [as illustrated in Fig.~\ref{figmodel}(a)], and $t_s$ controls the nonrelativistic spin splitting. The last term is the exchange coupling to N\'{e}el order with $J$ being the coupling strength, and we shall take $\bm N=\hat z$. For the SOC part, we can separately consider the spin-conserve ($\mathcal{H}_{\text{SOC}}^c$) and spin-flip ($\mathcal{H}_{\text{SOC}}^f$) components, with
\begin{eqnarray}
    \mathcal{H}_{\text{SOC}}^c &=& 4\lambda_c\sin(k_x/2)\sin(k_y/2)\tau_y\sigma_z,  \\
    \mathcal{H}_{\text{SOC}}^f &=& 4\lambda_f \sin(k_x/2)\cos(k_y/2)\tau_y\sigma_x \nonumber\\
    & &\qquad +4\lambda_f \cos(k_x/2)\sin(k_y/2)\tau_y\sigma_y,
\end{eqnarray}
where $\lambda$'s are the corresponding SOC strengths.

Figure~\ref{figmodel}(b) shows the calculated band structure without SOC, displaying characteristic $d$-wave altermagnetic spin splitting on $\Gamma$-$X/Y$ and $M$-$X/Y$ paths. Turning on SOC opens local gaps in the band structure [Fig.~\ref{figmodel}(c)]. The crossing points at $\mu\sim0.2$ eV are opened by spin-conserve SOC, since these degeneracies are composed of states with the same spins. Meanwhile, the degeneracies at $\mu\sim0.09$ eV are of opposite spins, so they are gapped by spin-flip SOC.

The MOHE conductivity is evaluated from
\begin{eqnarray}
\label{eq4}
    \sigma_{bc}^{L_a}=-\frac{e^2}{\hbar}\tau \sum_{n} \int \frac{d \bm{k}}{(2 \pi)^{d}} f_{0}' \langle j_b^{L_a}\rangle_{n\bm k}\langle v_c\rangle_{n\bm k},
\end{eqnarray}
where $\tau$ is the scattering time, $d$ is dimension of the system, $f_0$ is the Fermi distribution, $\langle v_c\rangle_{nm}=\langle u_{n\bm k}|\hat{v}_b|u_{m\bm k}\rangle$ are velocity matrix elements with $\langle v_c\rangle_{n\bm k}$ being the diagonal entries,
$\langle j_b^{L_a}\rangle_{n\bm k}\equiv \text{Re}\sum_m \langle v_b\rangle_{nm}\langle L_a\rangle_{mn}$, and $\langle \bm L\rangle_{mn}=(ie\hbar^2/4\mu_B)\sum_{p\neq m,n}(\frac{1}{\varepsilon_{p\bm k}-\varepsilon_{m\bm k}}+\frac{1}{\varepsilon_{p\bm k}-\varepsilon_{n\bm k}})\langle \bm v\rangle_{mp}\times\langle \bm v\rangle_{pn}$ are the matrix elements of orbital angular momentum~\cite{Dai2020gfactor,Manchon2022OHE,Souza2023multipole,Cysne2022OHE,Mertig2024OHE}.
According to symmetry analysis, this system should support component $\sigma_{xx}^{L_z}$.
In Fig.~\ref{figmodel}(d), we plot $\sigma_{xx}^{L_z}$ as a function of spin-conserve SOC strength $\lambda_c$ for Fermi level at $-0.04$ eV, away from band degeneracies [see Fig.~\ref{figmodel}(c)]. One observes that MOHE indeed vanishes when SOC is absent, and its value increases linearly with $\lambda_c$. If we change the SOC to spin-flip type, then $\sigma_{xx}^{L_z}$ exhibits a quadratic increase, as shown in Fig.~\ref{figmodel}(e).
These results are consistent with our perturbative results in Table~\ref{tableSOC}.

More importantly, MOHE can acquire nonperturbative enhancement when the Fermi level lies close to a SOC-induced gap. In Fig.~\ref{figmodel}(f), we plot the result for a Fermi level at $0.06$ eV, near the small gap opened by spin-flip SOC. Clearly, the dependence on spin-flip SOC strength $\lambda_f$ deviates from the quadratic scaling predicted by perturbative analysis. Moreover, the magnitude of MOHE exhibits a sharp increase with $\lambda_f$, showing the significant enhancement from SOC gaps. From the formula (\ref{eq4}), this nonperturbative characteristic originates from interband coherence amplified by band near-degeneracies.


\begin{figure}
	\includegraphics[width=8.5cm]{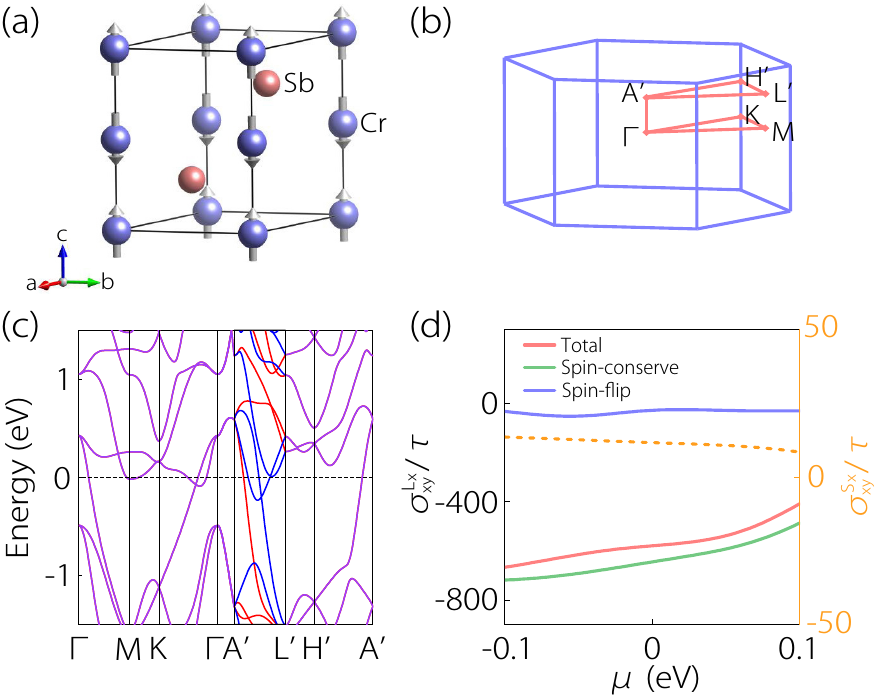}
	\caption{(a) Crystal structure, (b) Brillouin zone, and (c) nonrelativistic band structure of CrSb.  
   (d) Calculated collinearly-polarized responses $\sigma_{xy}^{L_x}/\tau$ (red solid line) and $\sigma_{xy}^{S_x}/\tau$ (orange dotted line) versus chemical potential. The spin-conserve (green solid line) and spin-flip (blue solid line) contributions to $\sigma_{xy}^{L_x}/\tau$ are also plotted.
		\label{figCrSb}}
\end{figure}

Below, we shall see that the features observed above are also validated in real altermagnetic materials. Particularly, although the MOHE needs SOC, its magnitude in altermagnets can be significant and can dominate over spin Hall response, contrary to previous expectations.

\textcolor{blue}{\textit{CrSb: Giant collinearly-polarized MOHE.}}
We first consider the experimentally verified altermagnetic metal CrSb~\cite{reimers2024direct,ding2024large,yang2025three,zhou2025CrSb}.
The crystal structure of CrSb has hexagonal space group $P6_{3}/mmc$ [Fig.~\ref{figCrSb}(a)]. Its N\'{e}el temperature is $\sim700$~K, with N\'{e}el vector along $z$. And the spin Laue group is ${}^{2}6/{}^{2}m^{2}m^{1}m$. Figure~\ref{figCrSb}(c) shows the band structure without SOC computed using first-principles methods (calculation details in Ref. \cite{supp}).

According to Table~\ref{tableSOC}, this material supports only one CPOC component, $\sigma_{xy}^{L_x}$. Figure~\ref{figCrSb}(d) plots $\sigma_{xy}^{L_x}/\tau$ as a function of Fermi level $\mu$ at $T=300$~K. The corresponding component $\sigma_{xy}^{S_x}$ of magnetic spin Hall conductivity is also plotted for comparison. One observes in a wide energy window, $\sigma_{xy}^{L_x}$ and $\sigma_{xy}^{S_x}$ exhibit opposite signs, and $\sigma_{xy}^{L_x}$ is about 50 times greater than its spin counterpart.
Interestingly, the orbital and spin responses here originate from different types of SOC: we explicitly show that the orbital conductivity is mainly from the spin-conserve SOC, whereas the spin conductivity primarily arises from spin-flip SOC (Fig. S2 in Supplemental Material), a feature consistent with the results in Table~\ref{tableSOC}.

\begin{figure*}
	\includegraphics[width=17cm]{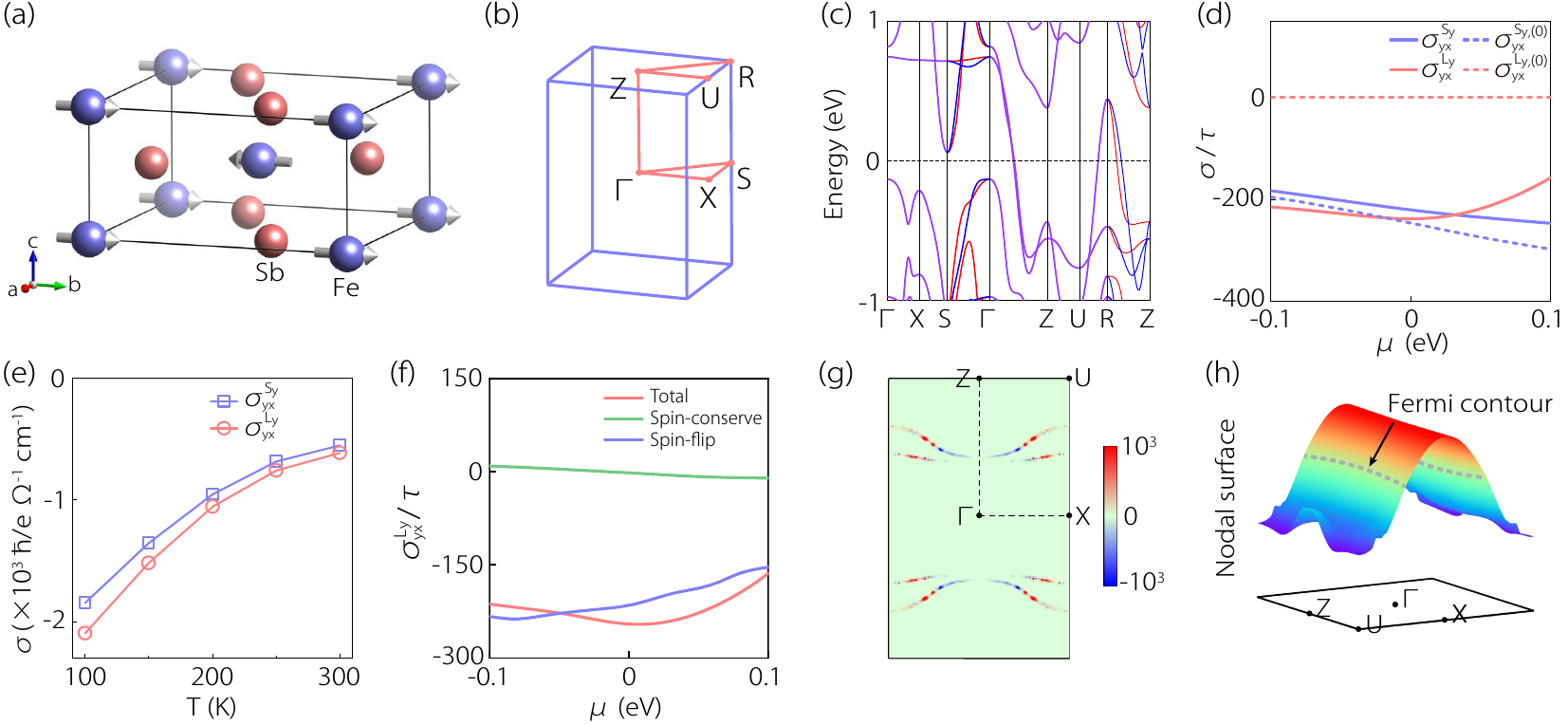}
	\caption{(a) Crystal structure, (b) Brillouin zone, and (c) non-relativistic band structure of FeSb$_2$. (d) Calculated $\sigma_{yx}^{L_y}/\tau$ and $\sigma_{yx}^{S_y}/\tau$ versus chemical potential. Here, $\sigma^{(0)}$'s (dashed lines) denote results in the absence of SOC. (e) Dependence of the results on temperature. The solids lines serve as a guide to the eyes. (f) Spin-conserve and spin-flip contributions to $\sigma^{L_y}_{yx}$. (g) Spin-flip contribution to the $k$-resolved $\sigma^{L_y}_{yx}$ (the integrand of Eq.~(\ref{eq4})) on the $k_y=0$ plane. (h) In the $k_y=0$ plane, 
two bands form a nodal surface in the absence of SOC. After SOC is turned on, the nodal surface is gapped, leading to the peak values in (g). 
		\label{figFeSb2}}
\end{figure*}

The scattering time can be estimated from experiment~\cite{peng2025scaling} as $\tau \sim 15$ fs at 300 K. Then, we find
$\sigma_{xy}^{L_x}$ reaches a giant value of $\sim -8703$~$(\hbar/e)\,\Omega^{-1}\mathrm{cm}^{-1}$ at 300 K [$\sigma_{xy}^{S_x}$ is $\sim 176$~$(\hbar/e)\,\Omega^{-1}\mathrm{cm}^{-1}$], and it can be further increased to be $\sim -3.1 \times 10^4$ ~$(\hbar/e)\,\Omega^{-1}\mathrm{cm}^{-1}$ at 100 K (Fig. S3 in Supplemental Material). For comparison, the previously reported collinearly-polarized spin Hall conductivities are usually less than $50$~$(\hbar/e)\,\Omega^{-1}\mathrm{cm}^{-1}$~\cite{Yang2024}, and a recent reported value of $100\sim200$~$(\hbar/e)\,\Omega^{-1}\mathrm{cm}^{-1}$ in  TaIrTe$_4$ was considered very large~\cite{Yang2023TaIrTe4,Yu2023TaIrTe4,Dash2024TaIrTe4}.
These demonstrate that the MOHE dominates the angular momentum transport in altermagnet CrSb, and it offers a remarkably large CPOC, desired for applications.

\textcolor{blue}{\textit{FeSb$_2$: Dominance over nonrelativistic spin response.}}
Our second example is altermagnetic FeSb$_2$~\cite{mazin2021prediction,attias2024intrinsic,phillips2025electronic}, which crystallizes in the orthorhombic structure with space group $Pnnm$.
Its Néel temperature is $\sim700$~K, with the easy axis along $y$ [Fig.~\ref{figFeSb2}(a)]. Figure~\ref{figFeSb2}(c) shows the calculated nonrelativistic band structure.
Differing from CrSb, the spin Laue group of FeSb$_2$ is ${}^{2}m^{2}m^{1}m$, which allows a nonrelativistic spin Hall conductivity $\sigma_{yx}^{S_y}$ in the absence of SOC, with the spin polarization parallel to the N\'eel vector. The symmetry of FeSb$_2$ also permits CPOC component $\sigma_{yx}^{L_y}$ in the same direction. Usually, one may guess that $\sigma_{yx}^{L_y}$ should be much less than $\sigma_{yx}^{S_y}$, given that $\sigma_{yx}^{L_y}$ is of higher order in SOC. Surprisingly, we find that the reality is not the case.


Figure~\ref{figFeSb2}(d) shows the calculated $\sigma_{yx}^{S_y}$ and $\sigma_{yx}^{L_y}$ in the absence and presence of SOC at $T=300$~K.
In the nonrelativistic limit, $\sigma_{yx}^{S_y}$ remains finite, whereas $\sigma_{yx}^{L_y}$ vanishes.
Upon introducing SOC, $\sigma_{yx}^{S_y}$ is not altered much, but $\sigma_{yx}^{L_y}$ is significantly increased, acquiring a magnitude comparable to $\sigma_{yx}^{S_y}$. Taking $\tau\sim 2.5$ fs as estimated from experiment~\cite{petrovic2003anisotropy}, we find $\sigma_{yx}^{L_y}\sim -613$~$(\hbar/e)\,\Omega^{-1}\mathrm{cm}^{-1}$ at 300 K, which is even larger than
$\sigma_{yx}^{S_y}\sim -552$~$(\hbar/e)\,\Omega^{-1}\mathrm{cm}^{-1}$. At lower temperatures, the responses become larger, and the orbital term remains greater than the spin one [Figure~\ref{figFeSb2}(e)].

This example clearly demonstrates that one has to be very careful in judging which effect is larger or weaker based solely on perturbative argument of the SOC dependence. Here, MOHE, despite being higher order in SOC, wins the competition against nonrelativistic spin Hall response.
%
To explore the origin of this counterintuitive result, we analyze the spin-conserve and spin-flip contributions to $\sigma_{yx}^{L_y}$ in Fig.~\ref{figFeSb2}(f), finding that the spin-flip SOC dominates the response.
In Fig.~\ref{figFeSb2}(g), we plot the spin-flip contribution to the $k$-resolved integrand in Eq.~(\ref{eq4}), showing notable peaks in the $k_y=0$ plane. This correlates with a dispersive nodal surface~\cite{Wu2018nodal} in the $k_y=0$ plane enforced by $[C_2\|\mathcal{M}_y]$ spin-group symmetry
and formed by two opposite spin states, as illustrated in Fig.~\ref{figFeSb2}(h). This nodal surface crosses the Fermi level with its SOC gap opened by spin-flip SOC, which
explains the nonperturbative enhancement of MOHE.

%

\textcolor{blue}{\textit{Discussion.}}
We have revealed that MOHE, while forbidden in conventional antiferromagnets, universally exists in the emerging class of altermagnetic materials. Compared to the previously highlighted nonrelativistic spin Hall effect~\cite{SST2021,Bose2022RuO2,Nitta2022,Song2023}, MOHE in altermagnets must require SOC. It has been a common practice in prior studies to infer relative importance of an effect in altermagnets based on its SOC dependence. However, this common wisdom fails here. We demonstrate concrete material examples where the MOHE reaches giant values, even surpassing the nonrelativistic spin response. Such enhancement is of nonperturbative character regarding SOC. It is connected to small SOC gaps near Fermi level, leading to increased interband coherence that contributes to orbital responses. The new understanding obtained here offers guideline for searching and designing orbital source materials with improved performance, and opens a new horizon for the application of altermagnets in information technology.

We have focused on MOHE which is $\mathcal{T}$-odd. There may also exist the conventional $\mathcal{T}$-even orbital Hall effect in altermagnets. In \cite{supp}, we also perform symmetry analysis of $\mathcal{T}$-even orbital (and spin) Hall effect in collinear altermagnets, and show that the CPOC from $\mathcal{T}$-even response is severely constrained, limited to only 4 spin Laue classes of altermagnets. In particular, CPOC from $\mathcal{T}$-even orbital Hall effect is forbidden in CrSb and FeSb$_2$.

Our work already predicts two concrete materials for exploring MOHE. To utilize MOHE for magnetic manipulation and switching, the generated CPOC current needs to be injected into an adjacent ferromagnetic layer, converted to spin current, and then drives the local moments in the ferromagnetic layer. This conversion process is described by the orbital-to-spin conversion efficiency $\eta$, which is a characteristic number for the ferromagnetic layer. For example, a large $\eta\sim 38\%$ can be achieved for Fe$_3$GaTe$_2$~\cite{zhang2025orbital}. One is usually concerning with the effective orbital Hall angle  $\theta^{*}_L\equiv(2e/\hbar)\eta\rho\sigma^{L}$ as a performance indicator, with $\rho$ being the resistivity of the orbital Hall material. We estimate that at room temperature, $\theta^{*}_L$ can reach $-37\%$ and $-15\%$ respectively for CrSb and FeSb$_2$, when Fe$_3$GaTe$_2$ is used as the ferromagnetic layer. Such values are much larger than the strongest reported spin Hall angles ($\sim 10\%$) with collinearly-polarized spin Hall current~\cite{Yang2023TaIrTe4,Yu2023TaIrTe4,Dash2024TaIrTe4,Loh2024}, suggesting them as good platforms for verifying our theory and for orbitronic applications.

\bibliography{ref}

@Misc{		  supp,
  howpublished	= "See Supplemental Material for theoretical and computational details."
}

@article{Wang2025Orbital-SRO,
  title={Unconventional scaling of the orbital Hall effect},
  author={Peng, Siyang and Zheng, Xuan and Li, Sheng and Lao, Bin and Han, Yamin and Liao, Zhaoliang and Zheng, Hongsheng and Yang, Yumeng and Yu, Tianye and Liu, Peitao and others},
  journal={Nature Materials},
  volume={24},
  number={11},
  pages={1749--1755},
  year={2025},
  publisher={Nature Publishing Group UK London},
  doi = {10.1038/s41563-025-02326-3},
  url = {https://doi.org/10.1038/s41563-025-02326-3}
}

@article{Yu2024SLC,
  title = {Predictable Gate-Field Control of Spin in Altermagnets with Spin-Layer Coupling},
  author = {Zhang, Run-Wu and Cui, Chaoxi and Li, Runze and Duan, Jingyi and Li, Lei and Yu, Zhi-Ming and Yao, Yugui},
  journal = {Phys. Rev. Lett.},
  volume = {133},
  issue = {5},
  pages = {056401},
  numpages = {6},
  year = {2024},
  month = {Aug},
  publisher = {American Physical Society},
  doi = {10.1103/PhysRevLett.133.056401},
  url = {https://link.aps.org/doi/10.1103/PhysRevLett.133.056401}
}

@article{Guo2014,
  title = {Anomalous Nernst and Hall effects in magnetized platinum and palladium},
  author = {Guo, G. Y. and Niu, Q. and Nagaosa, N.},
  journal = {Phys. Rev. B},
  volume = {89},
  issue = {21},
  pages = {214406},
  numpages = {6},
  year = {2014},
  month = {Jun},
  publisher = {American Physical Society},
  doi = {10.1103/PhysRevB.89.214406},
  url = {https://link.aps.org/doi/10.1103/PhysRevB.89.214406}
}

@article{2025Agterberg-quasisymmetry,
  title = {Quasisymmetry-Constrained Spin Ferromagnetism in Altermagnets},
  author = {Roig, Merc\`e and Yu, Yue and Ekman, Rune C. and Kreisel, Andreas and Andersen, Brian M. and Agterberg, Daniel F.},
  journal = {Phys. Rev. Lett.},
  volume = {135},
  issue = {1},
  pages = {016703},
  numpages = {8},
  year = {2025},
  month = {Jul},
  publisher = {American Physical Society},
  doi = {10.1103/839n-rckn},
  url = {https://link.aps.org/doi/10.1103/839n-rckn}
}

@article{Shao2023Neel,
  title = {N\'eel Spin Currents in Antiferromagnets},
  author = {Shao, Ding-Fu and Jiang, Yuan-Yuan and Ding, Jun and Zhang, Shu-Hui and Wang, Zi-An and Xiao, Rui-Chun and Gurung, Gautam and Lu, W. J. and Sun, Y. P. and Tsymbal, Evgeny Y.},
  journal = {Phys. Rev. Lett.},
  volume = {130},
  issue = {21},
  pages = {216702},
  numpages = {8},
  year = {2023},
  month = {May},
  publisher = {American Physical Society},
  doi = {10.1103/PhysRevLett.130.216702},
  url = {https://link.aps.org/doi/10.1103/PhysRevLett.130.216702}
}

@article{Bose2022RuO2,
  title={Tilted spin current generated by the collinear antiferromagnet ruthenium dioxide},
  author={Bose, Arnab and Schreiber, Nathaniel J and Jain, Rakshit and Shao, Ding-Fu and Nair, Hari P and Sun, Jiaxin and Zhang, Xiyue S and Muller, David A and Tsymbal, Evgeny Y and Schlom, Darrell G and others},
  journal={Nat. Electron.},
  volume={5},
  number={5},
  pages={267--274},
  year={2022},
  publisher={Nature Publishing Group UK London},
  doi = {10.1038/s41928-022-00744-8},
  url = {https://doi.org/10.1038/s41928-022-00744-8}
}

@article{Nitta2022,
  title = {Observation of Spin-Splitter Torque in Collinear Antiferromagnetic ${\mathrm{RuO}}_{2}$},
  author = {Karube, Shutaro and Tanaka, Takahiro and Sugawara, Daichi and Kadoguchi, Naohiro and Kohda, Makoto and Nitta, Junsaku},
  journal = {Phys. Rev. Lett.},
  volume = {129},
  issue = {13},
  pages = {137201},
  numpages = {6},
  year = {2022},
  month = {Sep},
  publisher = {American Physical Society},
  doi = {10.1103/PhysRevLett.129.137201},
  url = {https://link.aps.org/doi/10.1103/PhysRevLett.129.137201}
}

@article{Song2023,
  title = {Efficient Spin-to-Charge Conversion via Altermagnetic Spin Splitting Effect in Antiferromagnet ${\mathrm{RuO}}_{2}$},
  author = {Bai, H. and Zhang, Y. C. and Zhou, Y. J. and Chen, P. and Wan, C. H. and Han, L. and Zhu, W. X. and Liang, S. X. and Su, Y. C. and Han, X. F. and Pan, F. and Song, C.},
  journal = {Phys. Rev. Lett.},
  volume = {130},
  issue = {21},
  pages = {216701},
  numpages = {6},
  year = {2023},
  month = {May},
  publisher = {American Physical Society},
  doi = {10.1103/PhysRevLett.130.216701},
  url = {https://link.aps.org/doi/10.1103/PhysRevLett.130.216701}
}

@article{Mertig2024OHE,
  title = {Orbital Hall Effect Accompanying Quantum Hall Effect: Landau Levels Cause Orbital Polarized Edge Currents},
  author = {G\"obel, B\"orge and Mertig, Ingrid},
  journal = {Phys. Rev. Lett.},
  volume = {133},
  issue = {14},
  pages = {146301},
  numpages = {7},
  year = {2024},
  month = {Oct},
  publisher = {American Physical Society},
  doi = {10.1103/PhysRevLett.133.146301},
  url = {https://link.aps.org/doi/10.1103/PhysRevLett.133.146301}
}

@article{Yang2024,
  title={Field-free switching of perpendicular magnetization by two-dimensional PtTe2/WTe2 van der Waals heterostructures with high spin Hall conductivity},
  author={Wang, Fei and Shi, Guoyi and Kim, Kyoung-Whan and Park, Hyeon-Jong and Jang, Jae Gwang and Tan, Hui Ru and Lin, Ming and Liu, Yakun and Kim, Taeheon and Yang, Dongsheng and others},
  journal={Nat. Mater.},
  volume={23},
  number={6},
  pages={768--774},
  year={2024},
  publisher={Nature Publishing Group UK London},
  doi = {10.1038/s41563-023-01774-z},
  url = {https://doi.org/10.1038/s41563-023-01774-z}
}

@article{Loh2024,
  title={Two-dimensional chiral perovskites with large spin Hall angle and collinear spin Hall conductivity},
  author={Abdelwahab, Ibrahim and Kumar, Dushyant and Bian, Tieyuan and Zheng, Haining and Gao, Heng and Hu, Fanrui and McClelland, Arthur and Leng, Kai and Wilson, William L and Yin, Jun and others},
  journal={Science},
  volume={385},
  number={6706},
  pages={311--317},
  year={2024},
  publisher={American Association for the Advancement of Science},
  doi = {10.1126/science.adq0967},
  url = {https://doi.org/10.1126/science.adq0967}
}

@article{salemi2022theory,
  title={Theory of magnetic spin and orbital Hall and Nernst effects in bulk ferromagnets},
  author={Salemi, Leandro and Oppeneer, Peter M},
  journal={Phys. Rev. B},
  volume={106},
  number={2},
  pages={024410},
  year={2022},
  publisher={APS},
  doi = {10.1103/PhysRevB.106.024410},
  url = {https://link.aps.org/doi/10.1103/PhysRevB.106.024410}
}

@article{sala2022giant,
  title={Giant orbital Hall effect and orbital-to-spin conversion in 3 d, 5 d, and 4 f metallic heterostructures},
  author={Sala, Giacomo and Gambardella, Pietro},
  journal={Phys. Rev. Res.},
  volume={4},
  number={3},
  pages={033037},
  year={2022},
  publisher={APS},
  doi = {10.1103/PhysRevResearch.4.033037},
  url = {https://link.aps.org/doi/10.1103/PhysRevResearch.4.033037}
}

@article{Zheng2020SOT-Zr,
  title = {Magnetization switching driven by current-induced torque from weakly spin-orbit coupled Zr},
  author = {Zheng, Z. C. and Guo, Q. X. and Jo, D. and Go, D. and Wang, L. H. and Chen, H. C. and Yin, W. and Wang, X. M. and Yu, G. H. and He, W. and Lee, H.-W. and Teng, J. and Zhu, T.},
  journal = {Phys. Rev. Res.},
  volume = {2},
  issue = {1},
  pages = {013127},
  numpages = {8},
  year = {2020},
  month = {Feb},
  publisher = {American Physical Society},
  doi = {10.1103/PhysRevResearch.2.013127},
  url = {https://link.aps.org/doi/10.1103/PhysRevResearch.2.013127}
}

@article{Ding2020OT-CuO,
  title = {Harnessing Orbital-to-Spin Conversion of Interfacial Orbital Currents for Efficient Spin-Orbit Torques},
  author = {Ding, Shilei and Ross, Andrew and Go, Dongwook and Baldrati, Lorenzo and Ren, Zengyao and Freimuth, Frank and Becker, Sven and Kammerbauer, Fabian and Yang, Jinbo and Jakob, Gerhard and Mokrousov, Yuriy and Kl\"aui, Mathias},
  journal = {Phys. Rev. Lett.},
  volume = {125},
  issue = {17},
  pages = {177201},
  numpages = {6},
  year = {2020},
  month = {Oct},
  publisher = {American Physical Society},
  doi = {10.1103/PhysRevLett.125.177201},
  url = {https://link.aps.org/doi/10.1103/PhysRevLett.125.177201}
}

@article{go2020orbital,
  title={Orbital torque: Torque generation by orbital current injection},
  author={Go, Dongwook and Lee, Hyun-Woo},
  journal={Phys. Rev. Res.},
  volume={2},
  number={1},
  pages={013177},
  year={2020},
  publisher={APS},
  doi = {10.1103/PhysRevResearch.2.013177},
  url = {https://link.aps.org/doi/10.1103/PhysRevResearch.2.013177}
}

@article{zheng2024effective,
  title={Effective electrical manipulation of a topological antiferromagnet by orbital torques},
  author={Zheng, Zhenyi and Zeng, Tao and Zhao, Tieyang and Shi, Shu and Ren, Lizhu and Zhang, Tongtong and Jia, Lanxin and Gu, Youdi and Xiao, Rui and Zhou, Hengan and others},
  journal={Nat. Commun.},
  volume={15},
  number={1},
  pages={745},
  year={2024},
  publisher={Nature Publishing Group UK London},
  doi={10.1038/s41467-024-45109-1},
  url={https://doi.org/10.1038/s41467-024-45109-1}
}

@article{Otani2021OT,
  title = {Nontrivial torque generation by orbital angular momentum injection in ferromagnetic-metal/$\mathrm{Cu}/{\mathrm{Al}}_{2}{\mathrm{O}}_{3}$ trilayers},
  author = {Kim, Junyeon and Go, Dongwook and Tsai, Hanshen and Jo, Daegeun and Kondou, Kouta and Lee, Hyun-Woo and Otani, YoshiChika},
  journal = {Phys. Rev. B},
  volume = {103},
  issue = {2},
  pages = {L020407},
  numpages = {6},
  year = {2021},
  month = {Jan},
  publisher = {American Physical Society},
  doi = {10.1103/PhysRevB.103.L020407},
  url = {https://link.aps.org/doi/10.1103/PhysRevB.103.L020407}
}

@article{lee2021efficient,
  title={Efficient conversion of orbital Hall current to spin current for spin-orbit torque switching},
  author={Lee, Soogil and Kang, Min-Gu and Go, Dongwook and Kim, Dohyoung and Kang, Jun-Ho and Lee, Taekhyeon and Lee, Geun-Hee and Kang, Jaimin and Lee, Nyun Jong and Mokrousov, Yuriy and others},
  journal={Commun. Phys.},
  volume={4},
  number={1},
  pages={234},
  year={2021},
  publisher={Nature Publishing Group UK London},
  doi={10.1038/s42005-021-00737-7},
  url={https://doi.org/10.1038/s42005-021-00737-7}
}

@article{Jiang2024OT-PMA-Zr,
  title={Orbital torque switching in perpendicularly magnetized materials},
  author={Yang, Yuhe and Wang, Ping and Chen, Jiali and Zhang, Delin and Pan, Chang and Hu, Shuai and Wang, Ting and Yue, Wensi and Chen, Cheng and Jiang, Wei and others},
  journal={Nat. Commun.},
  volume={15},
  number={1},
  pages={8645},
  year={2024},
  publisher={Nature Publishing Group UK London},
  doi={10.1038/s41467-024-52824-2},
  url={https://doi.org/10.1038/s41467-024-52824-2}
}

@article{Klaui2025OT-PMA-Ru,
  title={Harnessing orbital Hall effect in spin-orbit torque MRAM},
  author={Gupta, Rahul and Bouard, Chlo{\'e} and Kammerbauer, Fabian and Ledesma-Martin, J Omar and Bose, Arnab and Kononenko, Iryna and Martin, Sylvain and Us{\'e}, Perrine and Jakob, Gerhard and Drouard, Marc and others},
  journal={Nat. Commun.},
  volume={16},
  number={1},
  pages={130},
  year={2025},
  publisher={Nature Publishing Group UK London},
  doi={10.1038/s41467-024-55437-x},
  url={https://doi.org/10.1038/s41467-024-55437-x}
}

@article{Libor2022,
  title = {Beyond Conventional Ferromagnetism and Antiferromagnetism: A Phase with Nonrelativistic Spin and Crystal Rotation Symmetry},
  author = {Šmejkal, Libor and Sinova, Jairo and Jungwirth, Tomas},
  journal = {Phys. Rev. X},
  volume = {12},
  issue = {3},
  pages = {031042},
  numpages = {16},
  year = {2022},
  doi = {10.1103/PhysRevX.12.031042},
  url = {https://link.aps.org/doi/10.1103/PhysRevX.12.031042}
}

@article{Libor2022TMR,
  title = {Giant and Tunneling Magnetoresistance in Unconventional Collinear Antiferromagnets with Nonrelativistic Spin-Momentum Coupling},
  author = {\ifmmode \check{S}\else \v{S}\fi{}mejkal, Libor and Hellenes, Anna Birk and Gonz\'alez-Hern\'andez, Rafael and Sinova, Jairo and Jungwirth, Tomas},
  journal = {Phys. Rev. X},
  volume = {12},
  issue = {1},
  pages = {011028},
  numpages = {11},
  year = {2022},
  month = {Feb},
  publisher = {American Physical Society},
  doi = {10.1103/PhysRevX.12.011028},
  url = {https://link.aps.org/doi/10.1103/PhysRevX.12.011028}
}

@article{peng2025scaling,
  title = {Scaling behavior of magnetoresistance and Hall resistivity in the altermagnet CrSb},
  author = {Peng, Xin and Wang, Yuzhi and Zhang, Shengnan and Zhou, Yi and Sun, Yuran and Su, Yahui and Wu, Chunxiang and Zhou, Tingyu and Liu, Le and Wang, Hangdong and Yang, Jinhu and Chen, Bin and Fang, Zhong and Du, Jianhua and Jiao, Zhiwei and Wu, Quansheng and Fang, Minghu},
  journal = {Phys. Rev. B},
  volume = {111},
  issue = {14},
  pages = {144402},
  numpages = {8},
  year = {2025},
  month = {Apr},
  publisher = {American Physical Society},
  doi = {10.1103/PhysRevB.111.144402},
  url = {https://link.aps.org/doi/10.1103/PhysRevB.111.144402}
}

@article{petrovic2003anisotropy,
  title = {Anisotropy and large magnetoresistance in the narrow-gap semiconductor ${\mathrm{FeSb}}_{2}$},
  author = {Petrovic, C. and Kim, J. W. and Bud'ko, S. L. and Goldman, A. I. and Canfield, P. C. and Choe, W. and Miller, G. J.},
  journal = {Phys. Rev. B},
  volume = {67},
  issue = {15},
  pages = {155205},
  numpages = {4},
  year = {2003},
  month = {Apr},
  publisher = {American Physical Society},
  doi = {10.1103/PhysRevB.67.155205},
  url = {https://link.aps.org/doi/10.1103/PhysRevB.67.155205}
}

@article{reimers2024direct,
  title={Direct observation of altermagnetic band splitting in CrSb thin films},
  author={Reimers, Sonka and Odenbreit, Lukas and {\v{S}}mejkal, Libor and Strocov, Vladimir N and Constantinou, Procopios and Hellenes, Anna B and Jaeschke Ubiergo, Rodrigo and Campos, Warlley H and Bharadwaj, Venkata K and Chakraborty, Atasi and others},
  journal={Nat. Commun.},
  volume={15},
  number={1},
  pages={2116},
  year={2024},
  publisher={Nature Publishing Group UK London},
  url = {https://doi.org/10.1038/s41467-024-46476-5}
}

@article{ding2024large,
  title={Large band splitting in g-wave altermagnet CrSb},
  author={Ding, Jianyang and Jiang, Zhicheng and Chen, Xiuhua and Tao, Zicheng and Liu, Zhengtai and Li, Tongrui and Liu, Jishan and Sun, Jianping and Cheng, Jinguang and Liu, Jiayu and others},
  journal={Phys. Rev. Lett.},
  volume={133},
  number={20},
  pages={206401},
  year={2024},
  publisher={APS},
  doi = {10.1103/PhysRevLett.133.206401},
  url = {https://link.aps.org/doi/10.1103/PhysRevLett.133.206401}
}

@article{yang2025three,
  title={Three-dimensional mapping of the altermagnetic spin splitting in CrSb},
  author={Yang, Guowei and Li, Zhanghuan and Yang, Sai and Li, Jiyuan and Zheng, Hao and Zhu, Weifan and Pan, Ze and Xu, Yifu and Cao, Saizheng and Zhao, Wenxuan and others},
  journal={Nat. Commun.},
  volume={16},
  number={1},
  pages={1442},
  year={2025},
  publisher={Nature Publishing Group UK London},
  url={https://doi.org/10.1038/s41467-025-56647-7},
  doi={10.1038/s41467-025-56647-7}
}

@article{mazin2021prediction,
  title={Prediction of unconventional magnetism in doped FeSb2},
  author={Mazin, Igor I and Koepernik, Klaus and Johannes, Michelle D and Gonz{\'a}lez-Hern{\'a}ndez, Rafael and {\v{S}}mejkal, Libor},
  journal={Proc. Natl. Acad. Sci. U.S.A.},
  volume={118},
  number={42},
  pages={e2108924118},
  year={2021},
  publisher={National Academy of Sciences},
  url = {https://doi.org/10.1073/pnas.2108924118}
}

@article{phillips2025electronic,
  title={Electronic structure of the altermagnet candidate FeSb 2: High-field torque magnetometry and density functional theory studies},
  author={Phillips, Cole and Pokharel, Ganesh and Shtefiienko, Kyryl and Bhandari, Shalika R and Graf, David E and Rai, DP and Shrestha, Keshav},
  journal={Phys. Rev. B},
  volume={111},
  number={7},
  pages={075141},
  year={2025},
  publisher={APS},
  doi = {10.1103/PhysRevB.111.075141},
  url = {https://link.aps.org/doi/10.1103/PhysRevB.111.075141}
}

@article{attias2024intrinsic,
  title={Intrinsic anomalous Hall effect in altermagnets},
  author={Attias, Lotan and Levchenko, Alex and Khodas, Maxim},
  journal={Phys. Rev. B},
  volume={110},
  number={9},
  pages={094425},
  year={2024},
  publisher={APS},
  doi = {10.1103/PhysRevB.110.094425},
  url = {https://link.aps.org/doi/10.1103/PhysRevB.110.094425}
}

@article{SST2021,
  title = {Efficient Electrical Spin Splitter Based on Nonrelativistic Collinear Antiferromagnetism},
  author = {Gonz\'alez-Hern\'andez, Rafael and \ifmmode \check{S}\else \v{S}\fi{}mejkal, Libor and V\'yborn\'y, Karel and Yahagi, Yuta and Sinova, Jairo and Jungwirth, Tom\'a\ifmmode \check{s}\else \v{s}\fi{} and \ifmmode \check{Z}\else \v{Z}\fi{}elezn\'y, Jakub},
  journal = {Phys. Rev. Lett.},
  volume = {126},
  issue = {12},
  pages = {127701},
  numpages = {6},
  year = {2021},
  month = {Mar},
  publisher = {American Physical Society},
  doi = {10.1103/PhysRevLett.126.127701},
  url = {https://link.aps.org/doi/10.1103/PhysRevLett.126.127701}
}

@article{Liu2025PRX,
  title = {Multipolar Anisotropy in Anomalous Hall Effect from Spin-Group Symmetry Breaking},
  author = {Liu, Zheng and Wei, Mengjie and Peng, Wenzhi and Hou, Dazhi and Gao, Yang and Niu, Qian},
  journal = {Phys. Rev. X},
  volume = {15},
  issue = {3},
  pages = {031006},
  numpages = {23},
  year = {2025},
  month = {Jul},
  publisher = {American Physical Society},
  doi = {10.1103/PhysRevX.15.031006},
  url = {https://link.aps.org/doi/10.1103/PhysRevX.15.031006}
}

@article{zhang2011PRL,
   author = {Zhang, H. and Freimuth, F. and Blugel, S. and Mokrousov, Y. and Souza, I.},
   title = {Role of spin-flip transitions in the anomalous Hall effect of FePt alloy},
   journal = {Phys. Rev. Lett.},
   volume = {106},
   number = {11},
   pages = {117202},
   DOI = {10.1103/PhysRevLett.106.117202},
   year = {2011},
   type = {Journal Article}
}

@article{model2024PRB,
  title = {Minimal models for altermagnetism},
  author = {Roig, Merc\`e and Kreisel, Andreas and Yu, Yue and Andersen, Brian M. and Agterberg, Daniel F.},
  journal = {Phys. Rev. B},
  volume = {110},
  issue = {14},
  pages = {144412},
  numpages = {20},
  year = {2024},
  month = {Oct},
  publisher = {American Physical Society},
  doi = {10.1103/PhysRevB.110.144412},
  url = {https://link.aps.org/doi/10.1103/PhysRevB.110.144412}
}

@article{antonenko2025mirror,
  title={Mirror Chern bands and Weyl nodal loops in altermagnets},
  author={Antonenko, Daniil S and Fernandes, Rafael M and Venderbos, J{\"o}rn WF},
  journal={Phys. Rev. Lett.},
  volume={134},
  number={9},
  pages={096703},
  year={2025},
  publisher={APS},
  doi = {10.1103/PhysRevLett.134.096703},
  url = {https://link.aps.org/doi/10.1103/PhysRevLett.134.096703}
}

@article{zhou2025CrSb,
   author = {Zhou, Z. and Cheng, X. and Hu, M. and Chu, R. and Bai, H. and Han, L. and Liu, J. and Pan, F. and Song, C.},
   title = {Manipulation of the altermagnetic order in CrSb via crystal symmetry},
   journal = {Nature},
   volume = {638},
   number = {8051},
   pages = {645-650},
   DOI = {10.1038/s41586-024-08436-3},
   url = {https://doi.org/10.1038/s41586-024-08436-3},
   year = {2025},
   type = {Journal Article}
}

@article{Ding2024orbital,
  title = {Orbital Torque in Rare-Earth Transition-Metal Ferrimagnets},
  author = {Ding, Shilei and Kang, Min-Gu and Legrand, William and Gambardella, Pietro},
  journal = {Phys. Rev. Lett.},
  volume = {132},
  issue = {23},
  pages = {236702},
  numpages = {7},
  year = {2024},
  month = {Jun},
  publisher = {American Physical Society},
  doi = {10.1103/PhysRevLett.132.236702},
  url = {https://link.aps.org/doi/10.1103/PhysRevLett.132.236702}
}

@article{zhang2025orbital,
  title={Orbital torque switching of room temperature two-dimensional van der Waals ferromagnet Fe3GaTe2},
  author={Zhang, Delin and Wei, Heshuang and Duan, Jinyu and Chen, Jiali and Chen, Jiaxin and Yue, Dongdong and Gong, Wanxi and Liu, Pengfei and Yang, Yuhe and Gou, Jinlong and others},
  journal={Nat. Commun.},
  volume={16},
  number={1},
  pages={7047},
  year={2025},
  publisher={Nature Publishing Group UK London},
  url={https://doi.org/10.1038/s41467-025-62333-5},
  doi={10.1038/s41467-025-62333-5}
}

@article{wu2007Fermi,
  title = {Fermi liquid instabilities in the spin channel},
  author = {Wu, Congjun and Sun, Kai and Fradkin, Eduardo and Zhang, Shou-Cheng},
  journal = {Phys. Rev. B},
  volume = {75},
  issue = {11},
  pages = {115103},
  numpages = {25},
  year = {2007},
  month = {Mar},
  publisher = {American Physical Society},
  doi = {10.1103/PhysRevB.75.115103},
  url = {https://link.aps.org/doi/10.1103/PhysRevB.75.115103}
}

@article{Liu2022spingroup,
  title = {Spin-Group Symmetry in Magnetic Materials with Negligible Spin-Orbit Coupling},
  author = {Liu, Pengfei and Li, Jiayu and Han, Jingzhi and Wan, Xiangang and Liu, Qihang},
  journal = {Phys. Rev. X},
  volume = {12},
  issue = {2},
  pages = {021016},
  numpages = {19},
  year = {2022},
  month = {Apr},
  publisher = {American Physical Society},
  doi = {10.1103/PhysRevX.12.021016},
  url = {https://link.aps.org/doi/10.1103/PhysRevX.12.021016}
}

@article{hayami2019momentum,
  title={Momentum-dependent spin splitting by collinear antiferromagnetic ordering},
  author={Hayami, Satoru and Yanagi, Yuki and Kusunose, Hiroaki},
  journal={j. Phys. Soc. Jpn.},
  volume={88},
  number={12},
  pages={123702},
  year={2019},
  publisher={The Physical Society of Japan},
  url = {https://doi.org/10.7566/JPSJ.88.123702},
  doi = {10.7566/JPSJ.88.123702}
}

@article{Yuan2020Giant,
  title = {Giant momentum-dependent spin splitting in centrosymmetric low-$Z$ antiferromagnets},
  author = {Yuan, Lin-Ding and Wang, Zhi and Luo, Jun-Wei and Rashba, Emmanuel I. and Zunger, Alex},
  journal = {Phys. Rev. B},
  volume = {102},
  issue = {1},
  pages = {014422},
  numpages = {13},
  year = {2020},
  month = {Jul},
  publisher = {American Physical Society},
  doi = {10.1103/PhysRevB.102.014422},
  url = {https://link.aps.org/doi/10.1103/PhysRevB.102.014422}
}

@article{Naka2019spincurrent,
   author = {Naka, Makoto and Hayami, Satoru and Kusunose, Hiroaki and Yanagi, Yuki and Motome, Yukitoshi and Seo, Hitoshi},
   title = {Spin current generation in organic antiferromagnets},
   journal = {Nat. Commun.},
   volume = {10},
   number = {1},
   pages = {4305},
   ISSN = {2041-1723},
   DOI = {10.1038/s41467-019-12229-y},
   url = {https://doi.org/10.1038/s41467-019-12229-y},
   year = {2019},
   type = {Journal Article}
}

@article{smejkal2020crystal,
   author = {Šmejkal, Libor and González-Hernández, Rafael and Jungwirth, T. and Sinova, J.},
   title = {Crystal time-reversal symmetry breaking and spontaneous Hall effect in collinear antiferromagnets},
   journal = {Sci. Adv.},
   volume = {6},
   number = {23},
   pages = {eaaz8809},
   DOI = {doi:10.1126/sciadv.aaz8809},
   url = {https://www.science.org/doi/abs/10.1126/sciadv.aaz8809},
   year = {2020},
   type = {Journal Article}
}

@article{ma2021multifunctional,
  title={Multifunctional antiferromagnetic materials with giant piezomagnetism and noncollinear spin current},
  author={Ma, Hai-Yang and Hu, Mengli and Li, Nana and Liu, Jianpeng and Yao, Wang and Jia, Jin-Feng and Liu, Junwei},
  journal={Nat. Commun.},
  volume={12},
  number={1},
  pages={2846},
  year={2021},
  publisher={Nature Publishing Group UK London},
  url={https://doi.org/10.1038/s41467-021-23127-7},
  doi={h10.1038/s41467-021-23127-7}
}

@article{song2025altermagnets,
   author = {Song, Cheng and Bai, Hua and Zhou, Zhiyuan and Han, Lei and Reichlova, Helena and Dil, J. Hugo and Liu, Junwei and Chen, Xianzhe and Pan, Feng},
   title = {Altermagnets as a new class of functional materials},
   journal = {Nat. Rev. Mater.},
   volume = {10},
   number = {6},
   pages = {473-485},
   ISSN = {2058-8437},
   DOI = {10.1038/s41578-025-00779-1},
   url = {https://doi.org/10.1038/s41578-025-00779-1},
   year = {2025},
   type = {Journal Article}
}

@article{smejkal2022prx,
  title = {Emerging Research Landscape of Altermagnetism},
  author = {\ifmmode \check{S}\else \v{S}\fi{}mejkal, Libor and Sinova, Jairo and Jungwirth, Tomas},
  journal = {Phys. Rev. X},
  volume = {12},
  issue = {4},
  pages = {040501},
  numpages = {27},
  year = {2022},
  month = {Dec},
  publisher = {American Physical Society},
  doi = {10.1103/PhysRevX.12.040501},
  url = {https://link.aps.org/doi/10.1103/PhysRevX.12.040501}
}

@article{Bai2024altermagnetism,
   author = {Bai, Ling and Feng, Wanxiang and Liu, Siyuan and Šmejkal, Libor and Mokrousov, Yuriy and Yao, Yugui},
   title = {Altermagnetism: Exploring New Frontiers in Magnetism and Spintronics},
   journal = {Adv. Funct. Mater.},
   volume = {34},
   number = {49},
   pages = {2409327},
   ISSN = {1616-301X},
   DOI = {https://doi.org/10.1002/adfm.202409327},
   url = {https://advanced.onlinelibrary.wiley.com/doi/abs/10.1002/adfm.202409327},
   year = {2024},
   type = {Journal Article}
}

@article{Dai2020gfactor,
  title = {Topological metals induced by the Zeeman effect},
  author = {Sun, Song and Song, Zhida and Weng, Hongming and Dai, Xi},
  journal = {Phys. Rev. B},
  volume = {101},
  issue = {12},
  pages = {125118},
  numpages = {10},
  year = {2020},
  month = {Mar},
  publisher = {American Physical Society},
  doi = {10.1103/PhysRevB.101.125118},
  url = {https://link.aps.org/doi/10.1103/PhysRevB.101.125118}
}

@Article{Souza2023multipole,
	title={{Multipole theory of optical spatial dispersion in crystals}},
	author={Óscar Pozo Ocaña and Ivo Souza},
	journal={SciPost Phys.},
	volume={14},
	pages={118},
	year={2023},
	publisher={SciPost},
	doi={10.21468/SciPostPhys.14.5.118},
	url={https://scipost.org/10.21468/SciPostPhys.14.5.118}
}

@article{kurebayashi2017view,
  title={Going in the right direction},
  author={Kurebayashi, Hidekazu},
  journal={Nat. Phys.},
  volume={13},
  number={3},
  pages={209--210},
  year={2017},
  publisher={Nature Publishing Group UK London},
  doi = {10.1038/nphys3954},
  url = {https://doi.org/10.1038/nphys3954}
}

@article{Yang2023briefing,
  title={Field-free and unconventional switching of perpendicular magnetization at room temperature},
  author={Yang, Hyunsoo and Liu, Yakun},
  journal={Nat. Electron.},
  volume={6},
  pages={724--725},
  year={2023},
  publisher={Nature Publishing Group UK London},
  doi = {10.1038/s41928-023-01040-9},
  url = {https://doi.org/10.1038/s41928-023-01040-9}
}

@article{Yang2023TaIrTe4,
  title={Field-free switching of perpendicular magnetization at room temperature using out-of-plane spins from TaIrTe4},
  author={Liu, Yakun and Shi, Guoyi and Kumar, Dushyant and Kim, Taeheon and Shi, Shuyuan and Yang, Dongsheng and Zhang, Jiantian and Zhang, Chenhui and Wang, Fei and Yang, Shuhan and others},
  journal={Nat. Electron.},
  volume={6},
  number={10},
  pages={732--738},
  year={2023},
  publisher={Nature Publishing Group UK London},
  doi = {10.1038/s41928-023-01039-2},
  url = {https://doi.org/10.1038/s41928-023-01039-2}
}

@article{Yu2023TaIrTe4,
  title={Room temperature field-free switching of perpendicular magnetization through spin-orbit torque originating from low-symmetry type II Weyl semimetal},
  author={Zhang, Yu and Xu, Hongjun and Jia, Ke and Lan, Guibin and Huang, Zhiheng and He, Bin and He, Congli and Shao, Qiming and Wang, Yizhan and Zhao, Mingkun and others},
  journal={Sci. Adv.},
  volume={9},
  number={44},
  pages={eadg9819},
  year={2023},
  publisher={American Association for the Advancement of Science},
  doi = {10.1126/sciadv.adg9819},
  url = {https://doi.org/10.1126/sciadv.adg9819}
}

@article{Dash2024TaIrTe4,
  title={Large out-of-plane spin--orbit torque in topological Weyl semimetal TaIrTe4},
  author={Bainsla, Lakhan and Zhao, Bing and Behera, Nilamani and Hoque, Anamul Md and Sj{\"o}str{\"o}m, Lars and Martinelli, Anna and Abdel-Hafiez, Mahmoud and {\AA}kerman, Johan and Dash, Saroj P},
  journal={Nat. Commun.},
  volume={15},
  number={1},
  pages={4649},
  year={2024},
  publisher={Nature Publishing Group UK London},
  doi = {10.1038/s41467-024-48872-3},
  url = {https://doi.org/10.1038/s41467-024-48872-3}
}

@Article{	  shao2021,
  title		= {Spin-neutral currents for spintronics},
  author	= {Shao, Ding-Fu and Zhang, Shu-Hui and Li, Ming and Eom,
		  Chang-Beom and Tsymbal, Evgeny Y.},
  journal	= {Nat. Commun.},
  volume	= {12},
  pages		= {7061},
  year		= {2021},
  doi		= {10.1038/s41467-021-26915-3},
  url		= {https://doi.org/10.1038/s41467-021-26915-3}
}

@article{lee2021orbital,
  title={Orbital torque in magnetic bilayers},
  author={Lee, Dongjoon and Go, Dongwook and Park, Hyeon-Jong and Jeong, Wonmin and Ko, Hye-Won and Yun, Deokhyun and Jo, Daegeun and Lee, Soogil and Go, Gyungchoon and Oh, Jung Hyun and others},
  journal={Nat. Commun.},
  volume={12},
  number={1},
  pages={6710},
  year={2021},
  publisher={Nature Publishing Group UK London},
  doi = {10.1038/s41467-021-26650-9
},
  url = {https://doi.org/10.1038/s41467-021-26650-9}
}

@article{bernevig2005orbitronics,
  title={Orbitronics: The intrinsic orbital current in p-doped silicon},
  author={Bernevig, B Andrei and Hughes, Taylor L and Zhang, Shou-Cheng},
  journal={Phys. Rev. Lett.},
  volume={95},
  number={6},
  pages={066601},
  year={2005},
  publisher={APS},
  doi = {10.1103/PhysRevLett.95.066601},
  url = {https://link.aps.org/doi/10.1103/PhysRevLett.95.066601}
}

@article{Guo2005,
  title = {Ab initio Calculation of the Intrinsic Spin Hall Effect in Semiconductors},
  author = {Guo, G. Y. and Yao, Yugui and Niu, Qian},
  journal = {Phys. Rev. Lett.},
  volume = {94},
  issue = {22},
  pages = {226601},
  numpages = {4},
  year = {2005},
  month = {Jun},
  publisher = {American Physical Society},
  doi = {10.1103/PhysRevLett.94.226601},
  url = {https://link.aps.org/doi/10.1103/PhysRevLett.94.226601}
}

@article{kontani2009giant,
  title={Giant orbital Hall effect in transition metals: Origin of large spin and anomalous Hall effects},
  author={Kontani, Hiroshi and Tanaka, T and Hirashima, DS and Yamada, K and Inoue, J},
  journal={Phys. Rev. Lett.},
  volume={102},
  number={1},
  pages={016601},
  year={2009},
  publisher={APS},
  doi = {10.1103/PhysRevLett.102.016601},
  url = {https://link.aps.org/doi/10.1103/PhysRevLett.102.016601}
}

@article{go2018intrinsic,
  title={Intrinsic spin and orbital Hall effects from orbital texture},
  author={Go, Dongwook and Jo, Daegeun and Kim, Changyoung and Lee, Hyun-Woo},
  journal={Phys. Rev. Lett.},
  volume={121},
  number={8},
  pages={086602},
  year={2018},
  publisher={APS},
  doi = {10.1103/PhysRevLett.121.086602},
  url = {https://link.aps.org/doi/10.1103/PhysRevLett.121.086602}
}

@article{choi2023observation,
  title={Observation of the orbital Hall effect in a light metal Ti},
  author={Choi, Young-Gwan and Jo, Daegeun and Ko, Kyung-Hun and Go, Dongwook and Kim, Kyung-Han and Park, Hee Gyum and Kim, Changyoung and Min, Byoung-Chul and Choi, Gyung-Min and Lee, Hyun-Woo},
  journal={Nature},
  volume={619},
  number={7968},
  pages={52--56},
  year={2023},
  publisher={Nature Publishing Group UK London},
  doi={10.1038/s41586-023-06101-9},
  url={https://doi.org/10.1038/s41586-023-06101-9}
}

\end{document}